\documentclass[tightenlines,superscriptaddress,eqsecnum,floats,nofootinbib,showpacs]{revtex4-2}
\usepackage{amssymb}
\usepackage{stmaryrd}
\usepackage{amsmath}
\usepackage{amsfonts}
\usepackage{mathrsfs}
\usepackage{CJK}
\usepackage{amsmath,amssymb,amsfonts}
\usepackage[utf8]{inputenc}
\usepackage{graphicx}
\usepackage{subfigure}
\usepackage{color} 
\def\be{\begin{equation}}
	\def\ee{\end{equation}}
\def\ba{\begin{eqnarray}}
	\def\ea{\end{eqnarray}}

\begin{document}

\title{Spinning Particle Dynamics and ISCO in Covariant Loop Quantum Gravity}
\author{Yongbin Du}
\affiliation{School of Physics and Astronomy, Sun Yat-sen University, Guangzhou, 510275, China}

\author{Yunlong Liu}
\affiliation{Department of Physics, South China University of Technology, Guangzhou 510641, China}

\author{Xiangdong Zhang\footnote{Corresponding author. scxdzhang@scut.edu.cn}}
\affiliation{Department of Physics, South China University of Technology, Guangzhou 510641, China}
\date{\today}

\begin{abstract}
	
In this paper, we investigate the motion of spinning particles in the background of covariant loop quantum gravity black holes, focusing on two distinct effective metric solutions. Both metrics incorporate a quantum parameter $\zeta$, which quantifies the corrections from loop quantum gravity. When $\zeta$ approaches zero, the spacetime reduces to the classical Schwarzschild solution. Using the pole-dipole approximation, we derive the equations of motion for spinning particles, accounting for the spin-curvature coupling. Our analysis reveals significant deviations in the behavior of the Innermost Stable Circular Orbit (ISCO) due to quantum effects. In the first effective metric, as $\zeta$ increases, the ISCO's radial position shifts, and for sufficiently large values of $\zeta$ (greater than 4.55), the ISCO disappears, allowing particles to hover above the black hole. In contrast, in the second metric, ISCOs persist even for large values of $\zeta$, albeit with a more restrictive spin range. These findings highlight the impact of loop quantum gravity corrections on the dynamics of spinning particles and provide insights into potential observational consequences for gravitational wave detections.

	\end{abstract}
	\maketitle

\section{introduction}
General relativity, as extensively discussed in both popular science literature and specialized textbooks, has profoundly transformed our understanding of the nature of spacetime. Yet, from a practical standpoint, it is equally significant for offering a method of studying celestial orbits that stands in stark contrast to Newtonian mechanics. This distinction is especially prominent in regions of intense gravitational fields, such as near black holes, where orbiting bodies with intrinsic spin follow trajectories markedly different from Newtonian predictions \cite{MPD}. Furthermore, the entire motion is often accompanied by the radiation of gravitational waves \cite{Abbott2016a, Abbott2016b, Abbott2017a, Abbott2017b, Abbott2017c}—a phenomenon completely absent in Newtonian gravity. The detection of gravitational waves by the LIGO and Virgo collaborations in 2016 marked a revolutionary advancement in astrophysics, providing direct evidence of binary black hole mergers \cite{Abbott2019, Abbott2021}. As two black holes inspiral towards each other, driven by the emission of gravitational radiation, they gradually lose orbital energy, moving closer until they reach the innermost stable circular orbit (ISCO) of the system. At this critical radius, the orbit transitions from stable to unstable, leading to a rapid plunge and ultimately, the merger of the black holes. This final stage is characterized by a burst of gravitational waves, which LIGO detected with remarkable precision.

Understanding the ISCO is therefore essential for interpreting the late inspiral phase of binary black hole mergers, as it dictates the point at which stable circular motion ceases, and the plunge phase begins. Since this concept entered the academic spotlight, many works have analyzed the ISCO in various spacetime configurations \cite{SuzukiMaeda}-\cite{Zaslavskii}. These models, in turn, allow for the extraction of physical parameters, such as mass and spin, from gravitational wave data, thus linking theoretical predictions to observable quantities. Building upon these works, Refs.~\cite{Jefremov2015}-\cite{Uchupol2016} considered a more realistic scenario: celestial bodies orbiting black holes possess intrinsic spin. The effects of spin on the trajectory and the radius of the ISCO are taken into account, which is one of the central concerns of this study. The most fundamental change brought about by spin is the deformation of the equation of motion. Specifically, the particle no longer travels along the geodesic of spacetime but instead deviates due to the coupling effect between the spacetime curvature and its spin. Typically, the equation is considered under the pole-dipole approximation, where the particle's 4-momentum and 4-velocity are no longer parallel. Consequently, for a timelike 4-momentum, the particle’s 4-velocity may become spacelike. Therefore, when considering the equation of motion, we must also incorporate the timelike constraint. In \cite{18liu}, the radius of ISCO for spinning particles in KN black hole has been studied, and it has been found to be smaller than that of non-spinning particles in the same spacetime \cite{Jefremov2015}.

However, despite General Relativity's undefeated record in facing rigorous experimental tests, there remains an underlying skepticism that it fully captures the essence of gravity. This doubt stems from its logical discord with the worldview of quantum theory. On a superficial level, the concept of a test particle's trajectory conflicts with the Heisenberg uncertainty principle. On a deeper level, the existence of an irrevocable one-way boundary—the event horizon of a black hole—challenges the unitarity of quantum theory in such backgrounds, destabilizing the very foundation of predictability in physics. This dilemma is epitomized by the black hole information paradox \cite{Hawking1976}-\cite{firewall}. Only a self-consistent and complete theory of quantum gravity could fundamentally reconcile these two experimentally robust yet seemingly incompatible theories.

Loop quantum gravity (LQG) is one such attempt to achieve this reconciliation \cite{ashtekar1986}-\cite{thiemann1998}. The core idea of this theory is to quantize the very geometry of spacetime, representing the gravitational field through discrete ``loops" or ``rings." Spacetime is conceptualized as having a discrete, network-like structure, implying that at extremely small scales (the Planck scale), it is not continuous but composed of fundamental units. From a semiclassical perspective, loop quantum gravity offers quantum corrections to some well-known black hole solutions in general relativity \cite{Modesto2008,Liu2022,Lin2024,Shi2024,Zhang:2023yps}. However, most of these solutions encounter the so-called "covariance" issue which is crucial for studying geodesics and perturbations \cite{Bolowald2015}. Since the solution even with small perturbations are usually no longer the solution of the same quantum corrected Hamiltonian constraint. To overcame this issue, recently, black hole solutions compatible with quantum theory, retaining the diffeomorphism invariance while employing the Hamiltonian formalism are constructed for a vacuum, spherically symmetric system \cite{zhangcong,Bojowald2024}. And hence addressing the covariance problem under this setup. This paper focuses on a semiclassical loop quantum gravity metric explored in \cite{zhangcong,Bojowald2024}. Based on this foundation, we will investigate the quantum gravitational effects on the orbital corrections of test particles orbiting black holes.

The structure of this paper is as follows: In Section 2, we briefly introduce two effective metrics derived from covariant loop quantum gravity, and provide the equations of motion for spinning particles under the pole-dipole approximation in these backgrounds, along with the corresponding timelike constraint. In Section 3, we focus on the first metric, analyze the effective potential for the motion of spinning particles in this spacetime, identify the ISCO, and discuss the characteristics of the ISCO parameters, comparing them with those in purely classical spacetime. In Section 4, we apply the same approach to analyze the ISCO in the background of the second metric. Finally, in Section 5, we summarize the results of this paper and offer a range of prospects for future work.

\section{The Equation of Motion of Spinning Particles}

\subsection{Black hole solution of covariant 4-Spacetime Theories}
In this subsection, we will provide a brief review of the black hole solutions in \cite{zhangcong}. To maintain diffeomorphism invariance, two different forms of the effective Hamiltonian that satisfy this condition lead to two corresponding metric solutions via the standard procedure \cite{standard}. The resulting 4-dimensional line element of a black hole in Boyer-Lindquist coordinates \((t, r, \theta, \phi)\) reads as
\begin{eqnarray}
ds^2 &=& -f[r] \, dt^2 + \frac{1}{f[r] h[r]} \, dr^2 + r^2 d\Omega^2
\end{eqnarray}
with solution 1:
\begin{eqnarray}\label{solution1}
f[r] &=& 1 - \frac{2M}{r} + \frac{\zeta^2 M^2}{r^2} \left(1 - \frac{2M}{r} \right)^2, \\
h[r] &=& 1,
\end{eqnarray}
and solution 2:
\begin{eqnarray}\label{solution2}
f[r] &=& 1 - \frac{2M}{r}, \\
h[r] &=& 1 + \frac{\zeta^2 M^2}{r^2} \left(1 - \frac{2M}{r} \right),
\end{eqnarray}
where \( M \) is the mass of the covariant black hole. The quantum parameter $\zeta = \sqrt{4\sqrt{3}\pi \gamma^3 \ell_p^2}/M$, with \( \gamma \) being the Barbero-Immirzi parameter and \( \ell_p \) the Planck length. Here, we use the normalized quantum parameter after loop quantization. 

Solution 1 is similar in form to a Reissner-Nordström (RN) black hole with a small charge. The equation $f(r) = 0$ has two positive roots, given by $r_{+} = 2M$ and $r_{-} = M^{2}\zeta^{2}/\beta - \beta/3$, where $\beta^{3} = 3M^{3}\zeta^{2}(\sqrt{81 + 3\zeta^{2}} - 9)$. Similar to the characteristics of an RN black hole, these two roots correspond to the outer horizon and inner horizon. The solution 2 features a transition surface $\mathcal{T}$ \cite{zhangcong,Bojowald2024}, which is a spacelike surface marking the boundary between the black hole and white hole regions. This is a common phenomenon in loop quantum gravity (LQG) black hole solutions, replacing the perplexing black hole singularity with a continuous and smooth spacetime transition region. Another intriguing aspect of this transition surface lies in its potential to address the black hole information paradox. Simply put, the information of matter falling into the black hole is not lost; instead, it traverses the transition surface and is eventually released in another spacetime region. Both solutions represent modifications of the Schwarzschild metric, and as $\zeta$ approaches zero, indicating minimal quantum gravitational effects, both solutions reduce to the classical vacuum spherically symmetric solution.

\subsection{Equation of motion of spinning particles}
To obtain a more convenient framework for studying the trajectory of particles in black hole spacetimes, we assume here that the particles in focus have masses much smaller than that of the black hole. Thus, they may be treated as test particles, devoid of any backreaction on the background spacetime. In principle, any test particle in spacetime should follow a geodesic path. However, if the particle possesses spin, its trajectory deviates, governed instead by the Mathisson-Papapetrou-Dixon (MPD) equations, which is derived under the pole-dipole approximation \cite{MPD,poledipole},
\begin{eqnarray}
\frac{Dp^{a}}{D\tau} &=& -\frac{1}{2} R^{a}_{\ bcd} v^{b} S^{cd}, \label{MPD1}\\
\frac{DS^{ab}}{D\tau} &=& -p^{b} v^{a} + p^{a} v^{b}\label{MPD2},
\end{eqnarray}
where $S^{cd}$ is the spin tensor that is antisymmetric, $R^{a}_{\ bcd}$ is the Riemann tensor, $p^{a}$ is the 4-momentum, and $v^{b}$ is the tangent 4-vector along the trajectory.$\tau$ serve as the affine parameter of the curve. A set of supplementary condition are also needed \cite{spinsupplement}
\begin{equation}\label{supplementary}
  S^{ab}p_{b}=0
\end{equation}
to eliminate the remaining undetermined degrees of freedom, which are related to the center of mass of a spinning particle and are observer-dependent. Thus the condition is not unique \cite{spinsupplement} and may yield different ISCO radius results \cite{differentISCO}. Using the spin tensor and 4-momentum, we can define the mass $m$ and spin $s$ of the particle as follows:
\begin{eqnarray}
p^{a} p_{a} &=& -m^2, \label{p^2}\\
S^{ab} S_{ab} &=& 2 m^2 s^2.
\end{eqnarray}
\eqref{p^2} means that $p^{a}$ is always timelike along the trajectory. 
The affine parameter can be normalized as
\begin{equation}\label{normalize}
  p^{a} v_{a}=-m.
\end{equation}
Hence, we could get the difference between 4-momentum and 4-velocity as follows \cite{u-v}:
\begin{eqnarray}
v^{a} - p^{a}/m = \frac{S^{a b} R_{b c d e} p^{c} S^{d e}}{(2 m^2 + \frac{1}{2} R_{h i j k} S^{h i} S^{j k})m}.\label{v-u}
\end{eqnarray}
We observe that the 4-velocity and 4-momentum of the test particle are not parallel, which is an another effects of spin.

For the sake of simplicity, we prioritize the motion in the equatorial plane $ (\theta = \pi/2) $ where $p^{\theta}$ and $S^{\mu \theta}=S^{\theta \mu}$ are vanishing. We introduce a spin vector to express the component of $S^{ab}$ 
\begin{equation}\label{spinvector}
  s^{a}=-\frac{1}{2 m^2}\epsilon^{a}_{bcd}p^{b}S^{cd},
\end{equation}
where $ \epsilon $ is the antisymmetric tensor with $ \epsilon_{0123} = 1 $. One find all the components of the spin vector are trivial except
\begin{equation}
  s^{\theta}=-s.
\end{equation}
Hence, the expression of the antisymmetric spin tensor $ S^{ab} $ can be re-described as
\begin{eqnarray}
S^{tr} &=& -S^{rt} = \sqrt{h[r]} \, r \, s \, p^{\phi}, \label{Str}\\
S^{t\phi} &=& -S^{\phi t} = -\frac{1}{f[r] \sqrt{h[r]}} \, s \, p^{r}, \label{Stphi}\\
S^{r\phi} &=& -S^{\phi r} = -f[r] \sqrt{h[r]} \frac{s}{r} p^{t}.\label{Srphi}
\end{eqnarray}
Now we get the difference between the 4-momentum and 4-velocity as
\begin{eqnarray}
		\label{dtdtau}
		v^{t}&=&\frac{\Pi_1 }{\Pi} \hat{p}^{t},\\
		\label{drdtau}
		v^{r}&=&\frac{\Pi_2  }{\Pi} \hat{p}^{r},\\
		\label{dphidtau}
		v^{\phi}&=&\frac{\Pi_3}{\Pi} \hat{p}^{\phi},
	\end{eqnarray}
where
\begin{align}
\Pi = 4r - s^2 \left( 2h[r] f'[r] + 2 f[r]^2 h'[r] \hat{p}^t{}^2 + (2r^2 h[r] f'[r]
- r^3 (2h[r] f''[r] + f'[r] h'[r])) \hat{p}^{\phi}{}^2 \right),
\end{align}
\begin{eqnarray}
\Pi_1 &=& 4r - 2s^2 (h[r] f'[r] + f[r] h'[r]), \\
\Pi_2 &=& 4r - 2s^2 h[r] f'[r], \\
\Pi_3 &=& 4r - rs^2 (f'[r] h'[r] + 2h[r] f''[r]).
\end{eqnarray}
For the spinning particles, the conserved quantities in this spacetime can be given as
\begin{equation}
C_{\xi} = p^a \xi_a + \frac{1}{2} S^{ab} \nabla_a \xi_b,
\end{equation}
where $\xi$ is the Killing vector. Note that the Killing vectors are $\xi^a = (\partial / \partial t)^a$ and $\phi^a = (\partial / \partial \phi)^a$.
From it, we acquire the energy $E$ and the angular momentum $J$ in the equatorial plane
\begin{eqnarray}
E &=& f[r] p^t + \frac{1}{2} r s \sqrt{h[r]} f'[r] p^{\phi}, \\
J &=& r^2 p^{\phi} + s f[r] \sqrt{h[r]} p^t.
\end{eqnarray}
We also define orbital angular momentum as $l=J-s$ which will be useful in the following section. By inverting the above relations and combining with \eqref{Str}-\eqref{Srphi}, we obtain
\begin{eqnarray}
p^t &=& \frac{2 r E - s \sqrt{h[r]} f'[r] J}{2 r f[r] - s^2 f[r] h[r] f'[r]}, \label{pt}\\
p^{\phi} &=& \frac{2 J - 2 s \sqrt{h[r]} E}{2 r^2 - r s^2 h[r] f'[r]}.\label{pphi}
\end{eqnarray}
It is worth noting that the missing term $p^r$ can be obtained by the normalization condition of the 4-momentum in \eqref{p^2}:
\begin{align}
(p^r)^2 = &\frac{h[r]}{\left( s^2 h[r] f'[r] - 2r \right)^2} 
\Bigg( 4 J s E \sqrt{h[r]} \left( 2 f[r] - r f'[r] \right)+ 4 E^2 \left( r^2 - s^2 f[r] h[r] \right) \nonumber \\
&\times  J^2 \left( s^2 h[r] f'[r]^2 - 4 f[r] \right) - f[r] \left( s^2 h[r] f'[r] - 2r \right)^2 \Bigg). \label{pr^2}
\end{align}
Without loss of generality, we use dimensionless parameters instead of the original parameters, as follows:
\begin{equation}
\tilde{E} = \frac{E}{m}, \quad \tilde{t} = \frac{t}{M}, \quad \tilde{r} = \frac{r}{M}, \quad \tilde{s} = \frac{s}{M}, \quad \tilde{J} = \frac{J}{m M}.
\end{equation}
For simplicity of expression, we omit the tilde on dimensionless parameters; for example, $J$ will represent $\tilde{J}$ in the following parts of the paper.

\subsection{Timelike condition of the orbits}

It can be seen in \eqref{p^2} that the momentum naturally follows the trajectory in a timelike manner. However,\eqref{v-u} reveals that the velocity does not always align with the momentum, and under the pole-dipole approximation \cite{poledipole} it may lead to some highly unphysical conclusions, for example, the particle's velocity exceeding the speed of light. To make sure the orbits exist physically, a superluminal constraint must be imposed on the velocity \cite{superluminal}, given by
\begin{equation}\label{timelikecondition}
v^a v_a = \frac{\dot{r}^2}{f[r] h[r]} - \dot{t}^2 f[r] + \dot{\phi}^2 g[r] < 0.
\end{equation}
The emergence of this superluminal issue arises due to the polar-dipole approximation in the construction of the MPD equations. If non-minimal spin-gravitational interactions are taken into account, as suggested by \cite{reference29, reference30, reference31, reference32}, the superluminal problem can be circumvented.

\section{The Motion of Spinning Particles in the solution 1 Spacetime}

\subsection{Effective potential}
\begin{figure}[h!]
    \centering
    \subfigure[$s = -1$, $l = 5$, $\zeta=8$]{
        \includegraphics[width=0.35\textwidth]{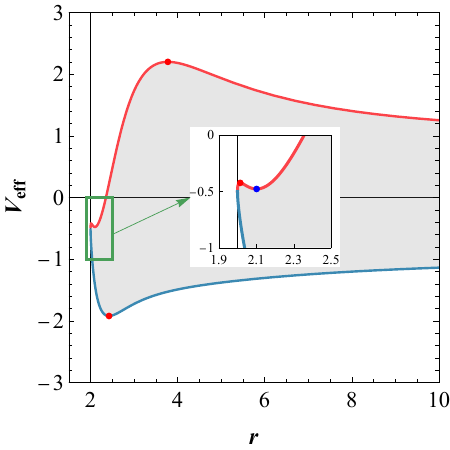}
    }
    \subfigure[$s = 1$, $l = 5$ and $\zeta=8$]{
        \includegraphics[width=0.35\textwidth]{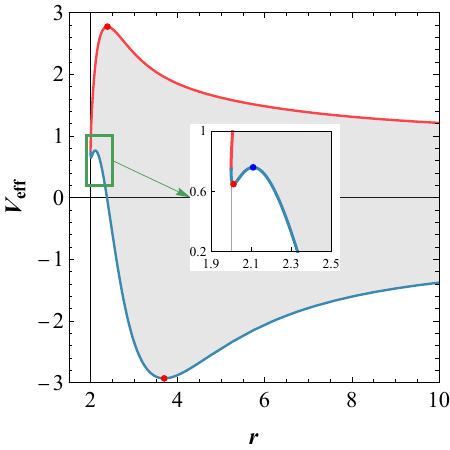}
    }
        \caption{Effective potential curves $V_{eff+}$ and $V_{eff-}$ as functions of the radial coordinate $r$. The red curve represents $V_{eff+}$, and the blue curve represents $V_{eff-}$. The shaded gray region indicates the forbidden zone for particle motion, where the condition $\mathcal{A}E^2 + \mathcal{B}E + \mathcal{C} < 0$ holds. Insets provide zoomed-in views of the regions near the extremal point, showing details of the effective potential behavior. $V_{eff-}$ can also attain positive values for some small radius region. The extremal point of the blue line in the right zoomed-in views represent a circular orbit as we will analyse in next subsection.}
    \label{Veff+-}
\end{figure}

The spherical symmetry of the spacetime background allows the orbital motion in the equatorial plane to be treated as a central force problem. As in Newtonian dynamics, where an effective potential is commonly used to study particle motion, an effective potential can similarly be introduced in general relativity to analyze the behavior of test particles in a fixed spacetime background. By writing the momentum \eqref{pr^2} in the following way as
\begin{align}
(\hat{p}^r)^2&=\mathcal{D} (\mathcal{A} E^2 + \mathcal{B} E + \mathcal{C}) \nonumber\\
&=\mathcal{D}(E-\frac{-\mathcal{B}+\sqrt{\mathcal{B}^2-4\mathcal{A}\mathcal{C}}}{2\mathcal{A}})(E-\frac{-\mathcal{B}-\sqrt{\mathcal{B}^2-4\mathcal{A}\mathcal{C}}}{2\mathcal{A}})
\end{align}
where
\begin{eqnarray}
\mathcal{A} &=& 4 \left( r^2 - s^2 f[r] h[r] \right), \\
\mathcal{B} &=& 4 J s \sqrt{h[r]} \left( 2 f[r] - r f'[r] \right), \\
\mathcal{C} &=& J^2 \left( s^2 h[r] f'[r]^2 - 4 f[r] \right) - f[r] \left( s^2 h[r] f'[r] - 2r \right)^2, \\
\mathcal{D} &=& \frac{h[r]}{\left( s^2 h[r] f'[r] - 2r \right)^2},
\end{eqnarray}
the effective potential of the background field for the particle can be defined as
\begin{eqnarray}
V_{eff+} &=& \frac{-\mathcal{B} + \sqrt{\mathcal{B}^2 - 4 \mathcal{A} \mathcal{C}}}{2 \mathcal{A}}, \\
V_{eff-} &=& \frac{-\mathcal{B} - \sqrt{\mathcal{B}^2 - 4 \mathcal{A} \mathcal{C}}}{2 \mathcal{A}}.
\end{eqnarray}
It has two branches. The negative branch $V_{eff-}$ is often neglected in typical ISCO effective potential studies due to its tendency to fall below zero \cite{18liu,24wei}. However, as shown in Fig.~\ref{Veff+-}, with appropriate parameter values(for example $s = 1$, $l = 5$ and $\zeta=8$ in the right plot), $V_{eff-}$ can also attain positive values for some small radius region. Thus we consider the negative branch when analyse the circular orbits. Since $(p^{r})^{2}$ is always greater than zero, the effective potential must satisfy $V_{eff+} \leq E$ or $V_{eff-} \geq E$ at any radial position $r$. It means that in the radial range where the particle actually moves, such as within $r_{1}$ to $r_{2}$, the total energy of the particle $E$ cannot lie within the shaded region bounded by the $V_{eff+}$ and $V_{eff-}$ curves in Fig~\ref{Veff+-}.

\begin{figure}[h!]
    \centering
    \subfigure[$\zeta = 1$, $l = 3.5$]{
        \includegraphics[width=0.2\textwidth]{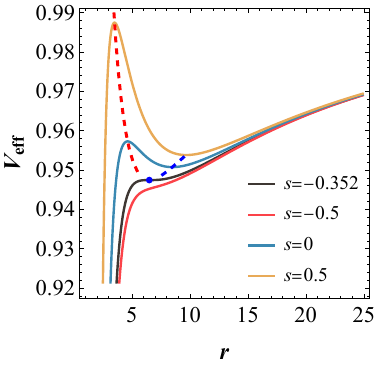}
    }
    \subfigure[$s = 0.5$, $l = 3.5$]{
        \includegraphics[width=0.2\textwidth]{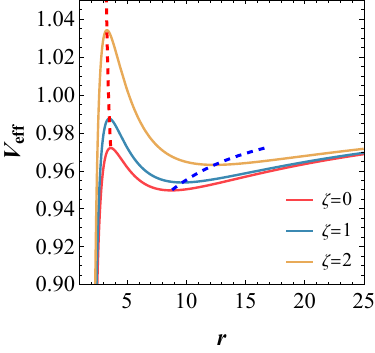}
    } 
    \subfigure[$s = 0.5$, $\zeta = 1$]{
        \includegraphics[width=0.2\textwidth]{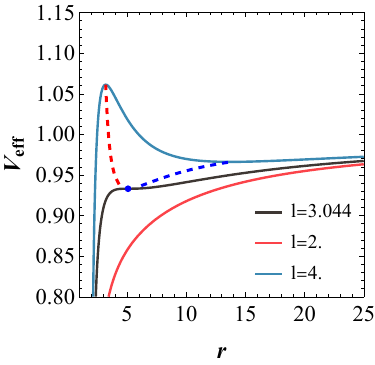}
    }
    \subfigure[$s = 0.5$, $\zeta = 1$]{
        \includegraphics[width=0.2\textwidth]{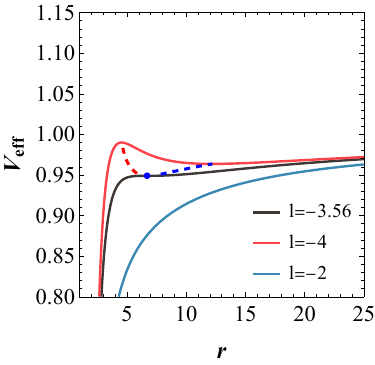}
    }
    \caption{The variation of $V_{eff+}$ with respect to the radial coordinate $r$. (a) Distinct $s$ for fixed $\zeta$ and $l$;  (b) Distinect $\zeta$ for fixed $s$ and $l$;  (c) Distinct positive $l$ for fixed $\zeta$ and $s$;  (d) Distinct negative $l$ for fixed $\zeta$ and $s$.}
    \label{G1_Veff_r}
\end{figure}

Now we focus on the variation of $V_{eff+}$ with respect to the radial coordinate $r$, as illustrated in Fig. \ref{G1_Veff_r}. Subplot (a) shows the effect of spin on $V_{eff}$, with fixed values $\zeta = 1$ and $l = 3.5$. As $s$ varies from $-0.352$ to $0.5$, the peak and shape of $V_{eff}$ undergo noticeable changes, demonstrating how spin can alter stability in the particle’s orbit. The smaller the spin, the lower the position of the corresponding curve. For the same curve, as the distance from the central force field increases, the effective potential first rises to a peak, then decreases, and finally turns upward again.
The overall trend of the curve in Subplot (b) is similar to that in (a), but it highlights the role of $\zeta$ in $V_{eff}$, where $s = 0.5$ and $l = 3.5$ are kept constant. The values of $\zeta$ range from $0$ to $2$, showing that the effect of LQG have on the effective potential profile. As $\zeta$ increases, the overall position of the curve shifts upward, and the peak becomes sharper. 
 Subplot (c) and (d) examines how changes in angular momentum parameter $l$, when taking positive and negative value, affect $V_{eff}$, holding $s = 0.5$ and $\zeta = 1$ fixed. It could be seen that as the absolute value of $l$ increases, the effective potential rises. When the absolute value of $l$ is relatively small, such as $l = 2$, the entire curve no longer exhibits roller-coaster-like fluctuations but instead rises smoothly.

\subsection{ISCO}
In this subsection we turn our attention to the Innermost Stable Circular Orbit. As the term ``circular'' implies, the trajectory of a particle in the ISCO is perfectly round, which necessitates that the radial velocity vanishes: $dr/d\tau = 0$. This criterion places a fundamental constraint on the effective potential, demanding that it satisfy 
\begin{equation}\label{V=E}
  V_{eff}=E.
\end{equation}
This condition also implies $d^{2}r/d\tau^{2} = 0$. Thus, the effective potential must also fulfill the following requirements:
\begin{equation}\label{dV}
  \frac{dV_{eff}}{dr}=0
\end{equation}
Beyond the requirements for zero radial velocity and acceleration, ``innermost" implies a boundary in stability: no further stable circular orbit exists closer to the central mass \cite{innermost}. Thus, the ISCO represents the critical point at which the effective potential reaches a delicate balance between a local maximum and minimum, intersecting precisely at this juncture. This condition can be expressed as:
\begin{equation}\label{ddV}
  \frac{d^{2}V_{eff}}{dr^{2}}=0
\end{equation}
Therefore, for a given $V_{eff}-r$ expression, the orbital position of the ISCO can be uniquely determined using the three conditions mentioned above.

\begin{figure}[h!]
    \centering
    \subfigure[$s = 1$]{
        \includegraphics[width=0.35\textwidth]{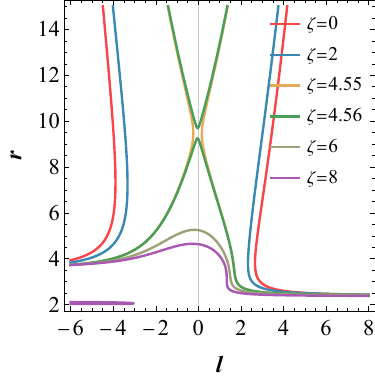} 
    }
     \subfigure[$\zeta = 4$]{
        \includegraphics[width=0.35\textwidth]{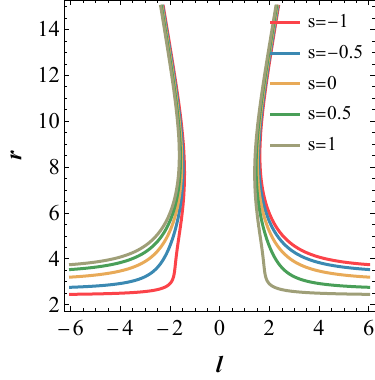} 
    }
    \caption{The radius of circular orbits as the function of $l$ when fixing $s$ (left panel) or $\zeta$ (right panel), for $V_{eff+}$ branch.}
    \label{r-l}
\end{figure}

\begin{figure}
  \centering
  \subfigure[$s=0.5$, $l=3.5$, $\zeta=1$, $E_{b} \approx 0.955$, $E_{g}=0.960$]{
        \includegraphics[width=0.3\textwidth]{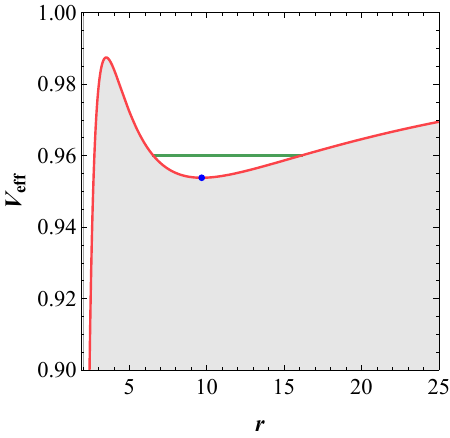} 
    }
     \subfigure[$s=3.6$, $l=1$, $\zeta=1$, $E_{b} \approx 0.673$, $E_{g}=0.680$]{
    	\includegraphics[width=0.3\textwidth]{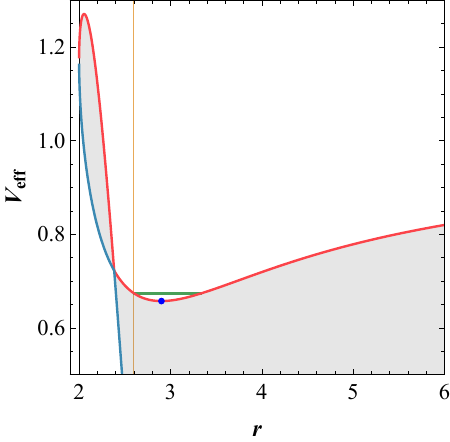} 
    }
    \subfigure[$s=1$, $l=0.0329$, $\zeta=4.6$, $E_{b} \approx 0.967$]{
    	\includegraphics[width=0.3\textwidth]{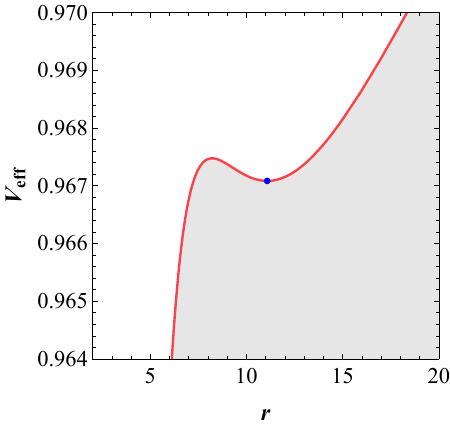} 
    }
    \subfigure[$s=0.5$, $l=3.5$, $\zeta=1$, $E_{b} \approx 0.955$, $E_{g}=0.960$]{
    	\includegraphics[width=0.3\textwidth]{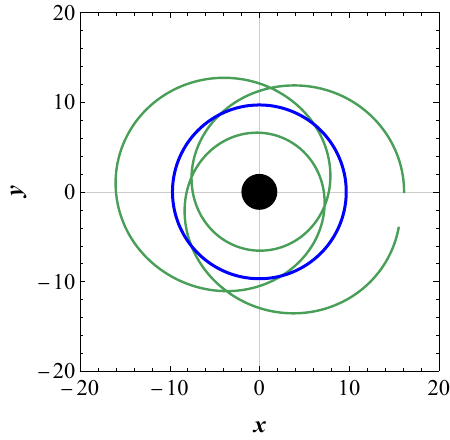} 
    }
    \subfigure[$s=3.6$, $l=1$, $\zeta=1$, $E_{b} \approx 0.673$, $E_{g}=0.680$]{
    	\includegraphics[width=0.3\textwidth]{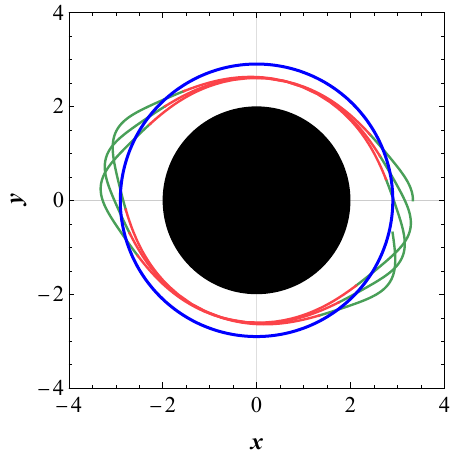} 
    }
    \subfigure[$s=1$, $l=0.0329$, $\zeta=4.6$, $E_{b} \approx 0.967$]{
        \includegraphics[width=0.3\textwidth]{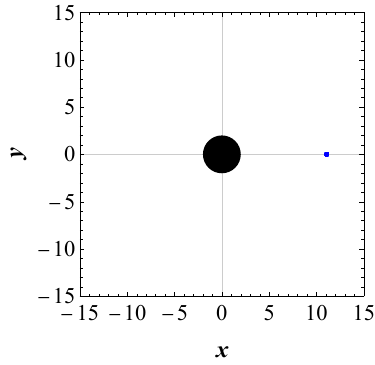} 
    }
  \caption{Subplot $(a)$-$(c)$ represent the $V_{eff+}-r$ curves under different values of $s$, $l$, and $\zeta$. The blue dots in the figures indicate the lowest points of the effective potential, corresponding to circular orbits with energy denoted by $E_{b}$. The green lines mark the potential positions of elliptical orbits with energy $E_{g}$. Subplot $(d)$-$(f)$ depict the detailed trajectories of orbits at energies $E_{b}$ and $E_{g}$, where the green elliptical trajectory may include spacelike regions, highlighted in red.}
  \label{circularorbit}
\end{figure}

As seen in Fig.~\ref{G1_Veff_r}, $V_{eff+}$ typically has two extremal points: one near the black hole, which is an unstable point, and another at a larger radius, which is stable. They both represent circular orbits, satisfying \eqref{V=E}. In Fig.~\ref{r-l}, we plot the variation of these two extremal points in the $r-l$ plane for different $\zeta$ or $s$. As shown in the left panel, when $s$ is set to $1$ and $\zeta$ is small, the curve resembles the results in \cite{18liu}, where the upper part of each turning point of the curve represents the stable extremal point's in Fig. \ref{G1_Veff_r}, while the lower part represents the unstable point. The turning points are also the tangent points of each curve with the vertical lines of constant $l$. The intersection of these two branches corresponds to the condition where the potential reaches a balance between a local maximum and minimum, which is precisely the point where \eqref{ddV} is satisfied. The ISCO lies at this intersection point. When $\zeta$ becomes sufficiently large, exceeding a certain critical value, the stable branch and unstable branch will no longer intersect, as observed in the left plot for $\zeta$ values greater than 4.56. In this case, the ISCO ceases to exist. This may be anticipated from Fig.~\ref{G1_Veff_r}, as the sharp peaks of the effective potential become increasingly steep with growing $\zeta$. This steepness makes it increasingly difficult for the maximum and minimum of the effective potential to coincide under appropriate spin conditions, ultimately leading to the disappearance of the ISCO. However, a very peculiar phenomenon may occur. As shown in (c) and (f) panel in Fig.~\ref{circularorbit} , when $\zeta > 4.56$ and $s=1$, we can find an specific angular momentum $l \approx 0.0329$ to obtain a solution, both the radial and angular velocity of which vanish, allowing the particle to hover at a position above the black hole! This strange phenomenon is precisely a loop quantum gravity effect, implying a repulsive force imposed on the celestial body, which is absent from the orbits in classical case \cite{18liu}.

The the right panel of Fig.~\ref{r-l} shows the effect of spin on the circular stable or unstable branches for fixed $\zeta$. When $\zeta=4$, an ISCO exists for all values of $s$ within the range $-1 \leq s \leq 1$. As $s$ increases, the angular momentum of ISCO decreases. The trend of its variation with spin is consistent with the results in \cite{18liu} where quantum effects are not considered.

\begin{figure}[h!]
    \centering
    \subfigure[$\zeta = 1$]{
        \includegraphics[width=0.22\textwidth]{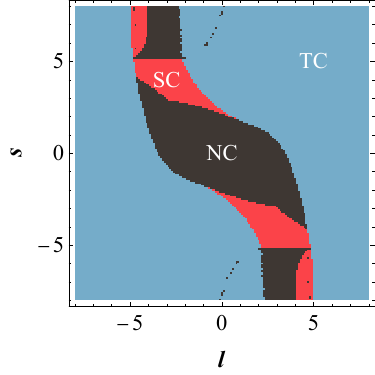} 
    }
     \subfigure[$\zeta = 3$]{
        \includegraphics[width=0.22\textwidth]{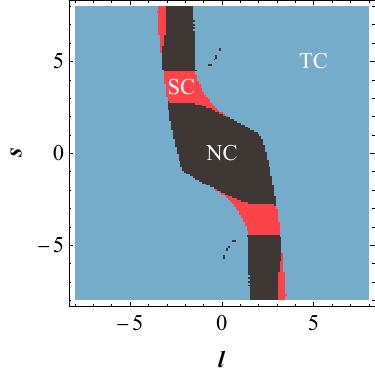} 
    }
      \subfigure[$\zeta = 4.55$]{
        \includegraphics[width=0.22\textwidth]{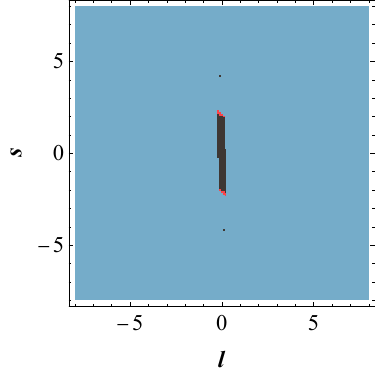} 
    }
   	 \subfigure[$\zeta = 1$]{
    	\includegraphics[width=0.22\textwidth]{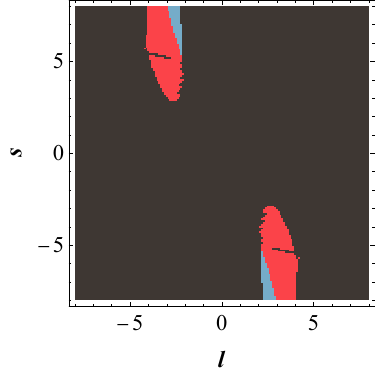} 
    }
    \caption{Allowed region of timelike circular orbits in $s-l$ plane. (a), (b) and (c) are for $V_{eff+}$ with different $\zeta$ and (d) is for $V_{eff-}$ with $\zeta=1$. The blue region, labeled as $TC$, represents the area where the timelike condition is satisfied, and circular orbits are allowed. The red region, labeled as $SC$, indicates the presence of circular orbits but with spacelike trajectories. The black region, labeled as $NC$, corresponds to areas where no circular orbits exist.}
    \label{timelike}
\end{figure}

When obtaining the ISCO through \eqref{V=E} to \eqref{ddV}, it is essential to verify whether the orbit is physical, i.e., whether it satisfies the timelike condition \eqref{timelikecondition}. In Fig.~\ref{timelike}, for a fixed $\zeta$, in the $s-l$ parameter space, regions satisfying the timelike condition and allowing circular orbits are represented in blue, while regions possess spacelike circular orbits are shown in red. Black regions do not has circular orbits. By comparing the left three panels, it is evident that as $\zeta$ increases, the parameter space for timelike circular orbits expands. The more pronounced the effects of loop quantum gravity, the smaller absolute value the angular momentum allowed for a timelike circular orbit. As a comparison, we can examine the $V_{eff-}$ branch. As illustrated in the Fig.~\ref{timelike} (d), it is evident that the blue region, representing timelike circular orbits, is small yet still present. 

To more clearly illustrate the characteristics of timelike circular orbits, in Fig.~\ref{circularorbit}, we present the rough shape of the orbit (see the three images in the second row, where the black circle represents the black hole, and the curves depict the particle trajectories) and the corresponding $V_{eff+}-r$ relation for a given parameter set of $s$, $l$, $\zeta$, $E_{g}$ and $E_{b}$, where $E_{g}$ is the total energy for orbits colored in green and $E_{b}$ for blue. 
In panel (a) and (d), the stable circular orbit, marked in blue dot and curve, associated with the $V_{eff+}$ branch occurs at approximately $r \approx 9.5$ with an energy of about $0.955$. The green trajectory whose energy is approximately 0.960, is elliptical. When taking $s=3.6$, $l=1$ and $\zeta=4.6$, the stable circular orbit is achieved at $E_{b} \approx 0.673$. However, the elliptical trajectory may exhibit spacelike regions at an energy of $E_{g} \approx 0.680$, which are marked in red in (e). The trajectory with both green and red segments is non-physical and needs to be excluded under the consideration of the timelike condition.
The panel (c) and (f) has been mentioned above when we discuss Fig.~\ref{r-l}. At this point, the angular velocity is vanishing the trajectory with energy $E_{b} \approx 0.9671$ hovers above the black hole, seeming to achieve a delicate balance between some repulsive force of loop quantum gravity and the classical gravitational attraction.

\begin{figure}[h!]
    \centering
    \subfigure[$\zeta = 1$]{
        \includegraphics[width=0.3\textwidth]{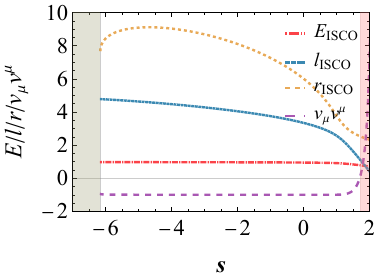} 
    }
    \subfigure[$\zeta = 2$]{
    	\includegraphics[width=0.3\textwidth]{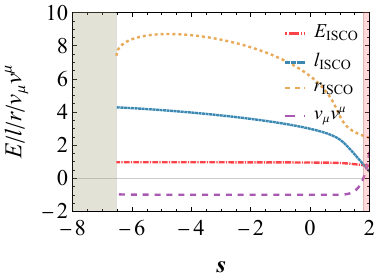} 
    }
     \subfigure[$\zeta = 3$]{
        \includegraphics[width=0.3\textwidth]{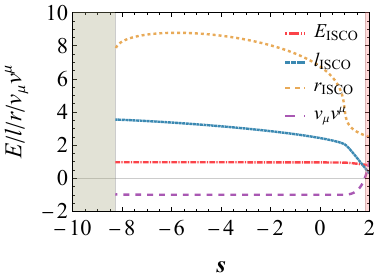} 
    }
    \caption{ISCO parameter $E_{ISCO},l_{ISCO},r_{ISCO}$ as the fuction of $s$ for (a) $\zeta=1$;  (b) $\zeta=2$;  (c) $\zeta=3$.}
    \label{ISCO}
\end{figure}

Finally we plot the ISCO parameter $E_{ISCO},l_{ISCO},r_{ISCO}$ as the fuction of $s$ in Fig.~\ref{ISCO} for different $\zeta$. The light pink region on the right of each panel represent spacelike region, where $v^{\mu}v_{\mu} > 0$ satisfy. The gray region does not contain an ISCO.  The energy at the ISCO, $E_{ISCO}$, depicted by the red dash-dotted line remains relatively stable across varying values of $s$ for different $\zeta$. The angular momentum $l_{ISCO}$, shown in blue dashed lines, decreases as $s$ approaches zero and then drops more steeply for positive values of $s$, meaning that the sign of spin would affect the orbital angular momentum for ISCO. The radius $r_{ISCO}$, plotted in orange dotted lines, initially rises with increasing $s$, reaching a peak, and then declines as $s$ further increases. It goes through rapid change, meaning that distinct $s$ give quite different circular orbits. 

To clearly show the effect of LQG on the allowed spin of ISCO, we also numerically solved for the spin values corresponding to the left gray boundary and the right pink boundary for different values of $\zeta$, and recorded them in Table \ref{G1sce}. 
The parameter $s_{c}$, which denotes the boundary spin value where the timelike orbit intersects with the edge of spacelike orbit, gradually increases with the growth of $\zeta$. In contrast, $s_{e}$, representing the minimum spin required for the ISCO orbit under a fixed loop quantum gravity-modified metric, does not exhibit a definitive trend in response to changes in $\zeta$. However, larger loop quantum gravity significantly gives larger allowable spin range for the ISCO. This influence could provide valuable insights in future observational experiments, such as gravitational wave detections, where the quantum aspects of gravity might be probed through phenomena associated with the ISCO.

	\begin{table}[!htb]
		\caption{\label{G1sce}
			The values of spins $s_c$ and $s_e$ corresponding to
			different LQG parameters $\zeta$. }
		\begin{ruledtabular}
			\begin{tabular}{lcll}
				$\zeta$ & $s_c$ & $s_e$				 \\
				0 & 1.652 & -6.15	\\
				1 & 1.706 & -6.16	\\
				2 & 1.798 & -6.52	\\
				3 & 1.814 & -8.27	
			\end{tabular}
		\end{ruledtabular}
	\end{table}
	

\section{The Motion of Spinning Particles in the solution 2 Spacetime}

\subsection{Effective potential}

\begin{figure}[h!]
    \centering
    \subfigure[$\zeta = 4$, $l = 3.5$]{
        \includegraphics[width=0.2\textwidth]{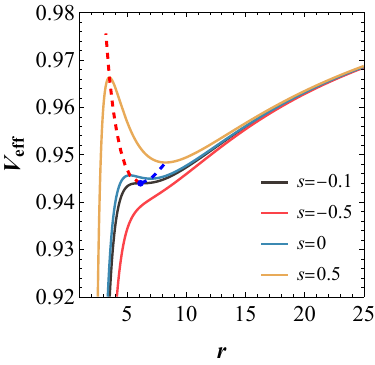}
    }
    \subfigure[$s = 0.5$, $l = 3.5$]{
    	\label{Veff2-r-2}
        \includegraphics[width=0.2\textwidth]{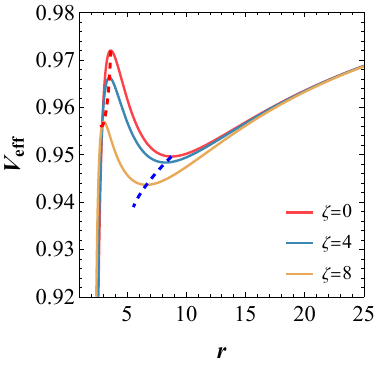}
    } 
    \subfigure[$s = 0.5$, $\zeta = 4$]{
        \includegraphics[width=0.2\textwidth]{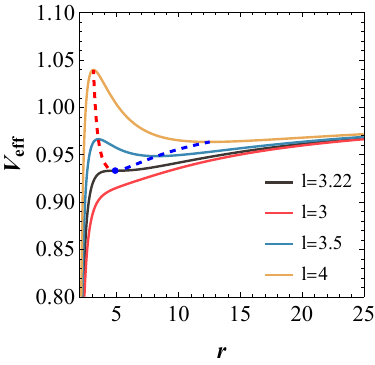}
    }
    \subfigure[$s = 0.5$, $\zeta = 4$]{
        \includegraphics[width=0.2\textwidth]{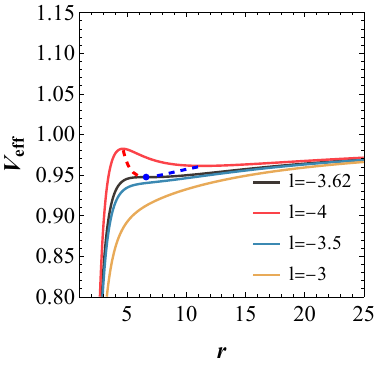}
    }
    \caption{The variation of $V_{eff+}$ with respect to the radial coordinate $r$. (a) Distinct $s$ for fixed $\zeta$ and $l$;  (b) Distinect $\zeta$ for fixed $s$ and $l$;  (c) Distinct positive $l$ for fixed $\zeta$ and $s$;  (d) Distinct negative $l$ for fixed $\zeta$ and $s$.}
    \label{Veff2-r}
\end{figure}

In this section, we apply a similar approach to investigate the effective potential and ISCO for the covariant gravity-corrected metric given by Eq.~\eqref{solution2}. We begin by examining the effective potential $V_{eff+}$, as illustrated in Fig.~\ref{Veff2-r}. The overall trend of the effective potential is similar to that of \eqref{solution1}. However, a notable difference appears in Fig.~\ref{Veff2-r-2}, where the effective potential diminishes as the loop quantum gravity effect strengthens, in contrast to the result from the first covariant solution. This implies that the two solutions are fundamentally different, and their respective $\zeta$ parameters have opposite effects on the ISCO trajectory, such as the permissible spin range. Our subsequent research has also confirmed this point.

\subsection{ISCO}
\begin{figure}[h!]
    \centering
    \subfigure[$s = 1$]{
        \includegraphics[width=0.3\textwidth]{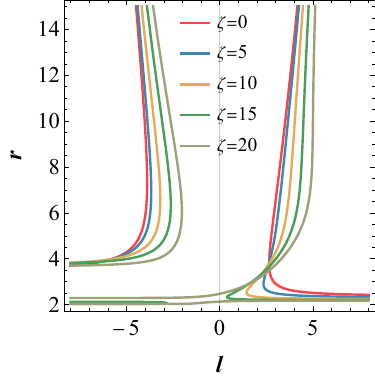}
    }
    \subfigure[$\zeta = 4$]{
        \includegraphics[width=0.3\textwidth]{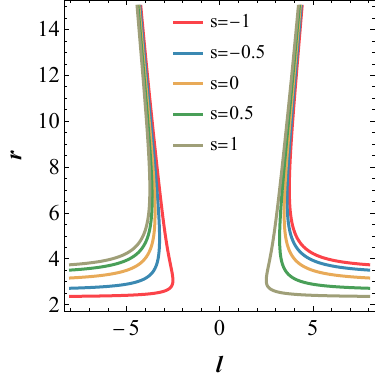}
    } 
    \caption{The radius of circular orbits as the function of $l$ when fixing $s$ (left panel) or $\zeta$ (right panel), for $V_{eff+}$ branch.}
    \label{r-l2}
\end{figure}

In Fig.~\ref{r-l2}, we plot the variation of these two extremal points in the $r - l$ plane for different values of $\zeta$ or $s$. As shown in the left panel of Fig.~\ref{r-l}, larger value of $\zeta$ may change the sign of the angular momentum of the turning point. A notable difference is observed in solution 2, where the ISCO remains present even when $\zeta$ is significantly large, contrasting sharply with solution 1. Once $\zeta$ is fixed, variations in spin consistently affect the turning point in the same way as solution 1: a higher spin results in a smaller ISCO angular momentum, as illustrated in the right panel. 

\begin{figure}[h!]
    \centering
    \subfigure[$\zeta = 4$]{
        \includegraphics[width=0.22\textwidth]{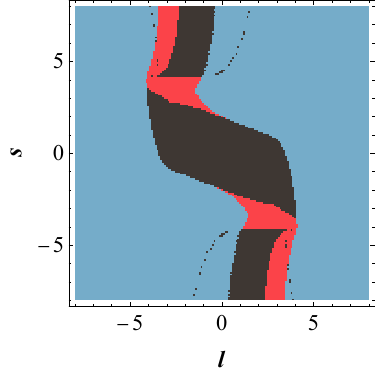}
    }
    \subfigure[$\zeta = 8$]{
        \includegraphics[width=0.22\textwidth]{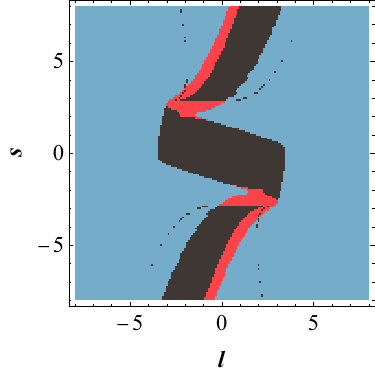}
    } 
     \subfigure[$\zeta = 16$]{
        \includegraphics[width=0.22\textwidth]{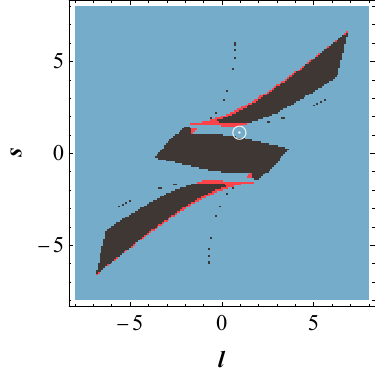}
    }
    \subfigure[$\zeta = 8$]{
    	\includegraphics[width=0.22\textwidth]{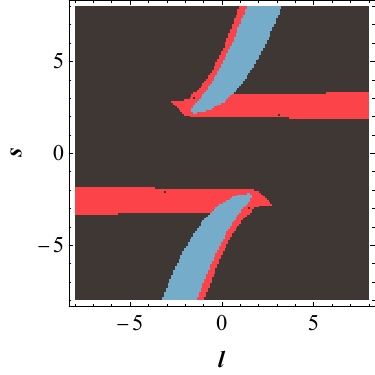}
    }
    \caption{Allowed region for timelike circular orbits in $s-l$ plane.}
    \label{timelike2}
\end{figure}

After considering the timelike condition, we illustrate the allowed regions for stable circular orbits in the $s-l$ plane for both $V_{eff+}$ and $V_{eff-}$ in Fig.~\ref{timelike2}. The blue-shaded regions still represent the parameter ranges where stable, timelike circular orbits exist. For the $V_{eff+}$ branch, as $\zeta$ increases from 4 to 16, we observe a relatively slight narrowing of the red and dark regions, representing spacetime circular orbits and no circular orbits respectively, indicating that the allowable parameter space for stable timelike circular orbits increases as the loop quantum gravity effects intensify. For $V_{eff-}$ branch, allowed region is quite small but still exist, and we find it getting larger as $\zeta$ getting bigger.

\begin{figure}[h!]
    \centering
    \subfigure[$\zeta = 2$]{
        \includegraphics[width=0.3\textwidth]{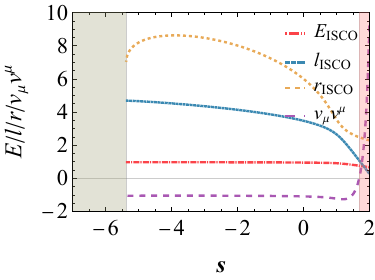}
    }
    \subfigure[$\zeta = 4$]{
        \includegraphics[width=0.3\textwidth]{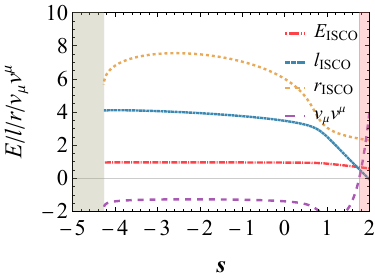}
    } 
     \subfigure[$\zeta = 8$]{
        \includegraphics[width=0.3\textwidth]{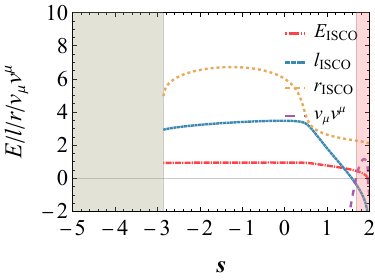}
    }
    \caption{ISCO parameter $E_{ISCO}$,$l_{ISCO}$ and $r_{ISCO}$ as a function of $s$ for (a) $\zeta=2$;  (b) $\zeta=4$;  (c) $\zeta=8$.}
    \label{ISCO2}
\end{figure}

Through the analysis above, we have finally plotted the variation of ISCO parameters as a function of spin in Fig.~\ref{ISCO2}. 
We also provide the numerical results for $s_{c}$ and $s_{e}$ under various values of $\zeta$ in Table \uppercase\expandafter{\romannumeral2}. It can be observed that $s_{c}$ changes gradually with increasing $\zeta$, whereas $s_{e}$ grows rapidly due to the pronounced effects of loop quantum gravity. Specifically, under the conditions of solution 2, stronger quantum effects impose stricter constraints on the timelike ISCO, thereby reducing the permissible spin range. This stands in stark contrast to the result observed in solution 1.
	\begin{table}[!htb]
		\caption{\label{G2sce}
			The values of spins $s_c$ and $s_e$ corresponding to
			different LQG parameters $\zeta$. }
		\begin{ruledtabular}
			\begin{tabular}{lcll}
				$\zeta$ & $s_c$ & $s_e$				 \\
				0M & 1.652 & -6.15	\\
				2M & 1.692 & -5.38	\\
				4M & 1.761 & -4.26	\\
				6M & 1.762 & -3.45	\\
				8M & 1.687 & -2.86 	
			\end{tabular}
		\end{ruledtabular}
	\end{table}


\section{Conclusion and Discussion}

In this paper, we have examined the dynamics of spinning particles in covariant loop quantum gravity black hole spacetimes, focusing on two effective metric solutions. 
Both of these metrics include a loop quantum characteristic parameter $\zeta$, and when it tends to zero, the spacetime background reduces to the Schwarzschild metric.
Due to the spin-curvature force, the particle's trajectory deviates from the geodesic, and its 4-momentum is no longer parallel to its 4-velocity.
By employing the pole-dipole approximation, we derived the modified equations of motion and imposed timelike constraints to ensure physically meaningful solutions. 
Using the expression for the radial momentum, we introduced the effective potential to simplify the analysis of the motion.
Based on the three conditions for the ISCO described by the effective potential, we derived the ISCO in the spacetime with loop quantum corrections and provided the corresponding parametric relationships.
Our analysis of the effective potential revealed significant deviations from classical general relativity, particularly in the behavior of the Innermost Stable Circular Orbit (ISCO) and particle trajectories.

For the first effective metric, we found that the overall height and steepness of the effective potential increase with the loop quantum gravity parameter $\zeta$, which causes a shift in the radial position of circular orbits compared to the Schwarzschild spacetime, thereby further affecting the radial location of the ISCO. 
Additionally, we found that for a given particle spin, such as $s=1$, if $\zeta$ exceeds approximately 4.55, the entire spacetime no longer possesses an ISCO, a feature distinct from the purely classical spacetime background. This also leads to solutions where particles may hover above the black hole, and we have illustrated the corresponding trajectories in this paper. 
Furthermore, due to the timelike constraint, a larger $\zeta$ imposes less stringent requirements on the spin of particles capable of forming an ISCO, which may provide guidance for observing quantum gravitational effects.

In contrast, the second effective metric displays a different behavior. 
The effective potential decreases and flattens as $\zeta$ increases. Perhaps for this reason, the ISCO persists even when $\zeta$ takes much larger values compared to solution 1. 
Moreover, larger values of $\zeta$ impose stricter constraints on the allowable spin range for the ISCO, which is in stark contrast to the behavior observed in solution 1. This prompts us to consider that, although both metrics represent quantum corrections to gravity under the aim of preserving covariance, their effects on the behavior of orbiting particles, particularly the ISCO, may differ significantly. This suggests that the two metrics possess notable differences in their spacetime structures. Why the same covariance principle leads to such distinct structures is a question we will continue to explore.

 Future studies could extend this analysis by exploring other black hole solutions, such as those with higher-dimensional spacetimes or alternative quantum gravity frameworks. Additionally, further investigation into the role of quantum effects in more realistic, rotating black hole spacetimes (Kerr or Kerr-Newman) may offer deeper insights into the interplay between spin and quantum geometry. Gravitational wave observations, such as those detected by LIGO, provide an exciting opportunity to test these predictions. Detecting deviations from classical predictions in the inspiral and merger phases of black hole binaries could offer a window into the quantum nature of gravity. Overall, our findings lay the groundwork for future studies to bridge the gap between quantum gravity and observational astrophysics, enhancing our understanding of black hole interiors and the quantum structure of spacetime.

\begin{acknowledgements}
This work is supported by National Natural Science Foundation of China (NSFC) with Grants No.12275087. 
\end{acknowledgements}

\appendix
\par
\bibliographystyle{unsrt}

\end{document}